\documentclass[pra,twocolumn]{revtex4-1}

\pdfoutput=1

\usepackage{amsmath}
\usepackage{amssymb}
\usepackage{amsfonts}

\usepackage[customcolors]{hf-tikz}

\usepackage{multirow}

\usepackage{graphicx}
\usepackage{epstopdf}

\usepackage{braket}
\def\KetBra#1#2{\Ket{#1}\!\!\Bra{#2}}

\usepackage{dsfont}

\graphicspath{{images/}}

\begin{document}

\title{Adiabatic quantum optimization in presence of discrete noise: Reducing the problem dimensionality}
\author{Salvatore Mandr\`a$^{*}$}
\author{Gian Giacomo Guerreschi$^{*}$}
\author{Al\'an Aspuru-Guzik}
\affiliation{Department of Chemistry and Chemical Biology, Harvard University,
  12 Oxford Street, 02138 Cambridge MA}

\begin{abstract}
Adiabatic quantum optimization is a procedure to solve a vast class of optimization
problems by slowly changing the Hamiltonian of a quantum system. The evolution time
necessary for the algorithm to be successful scales inversely with the minimum energy
gap encountered during the dynamics. Unfortunately, the direct calculation of the gap
is strongly limited by the exponential growth in the dimensionality of the Hilbert space
associated to the quantum system. Although many special-purpose methods have been
devised to reduce the effective dimensionality, they are strongly limited to particular
classes of problems with evident symmetries. Moreover, little is known about the
computational power of adiabatic quantum optimizers in real-world conditions.
Here, we propose and implement a general purposes reduction method that does not rely
on any explicit symmetry and which requires, under certain general conditions,
only a polynomial amount of classical resources. Thanks to this method, we are
able to analyze the performance of ``non-ideal'' quantum adiabatic optimizers
to solve the well-known Grover problem,
namely the search of target entries in an unsorted database, in the presence of discrete
local defects. In this case, we show that adiabatic quantum optimization, even if affected
by random noise, is still potentially faster than any classical algorithm.
\end{abstract}


\maketitle

\section{Introduction}
\label{sec:introduction}

In 2001, Farhi \emph{et al.} \cite{Farhi2001a} proposed a new paradigm to carry
out quantum computation (QC) that is based on the adiabatic evolution of a quantum system
under a slowly changing Hamiltonian and that builds on previous results developed by the
statistical and chemical physics communities in the context of quantum annealing techniques
\cite{Ray1989,Kadowaki1998,Finnila1994,Lee2000}.
While this approach constitutes an alternative framework in which the development of
new quantum algorithms for optimization problems results more intuitive
\cite{Babbush2012,somma2012quantum,perdomo2012finding},
the estimation of the evolution time and its scaling with the problem size still
remains unclear. For example, factors like the choice of the schedule
\cite{Roland2002,Farhi2002} and the specific form of the Hamiltonian \cite{Das2002,Wei2006,altshuler2010anderson,Choi2011}
influence the adiabatic evolution in ways that are, so far, not fully understood.
Adiabatic QC at zero temperature has been proved to be polynomially equivalent
to the usual QC with gates and circuits \cite{vanDam2001,Aharonov2008} and, therefore, any
exponential quantum speedup should be attainable \cite{Deutsch1992,Grover1996,Shor1999,Abrams1999}.
In this respect, several numerical studies
carried out in the last few years reported encouraging results for small systems
\cite{Farhi2001a,Hogg2003,Perdomo-Ortiz2010,Choi2011,hen2012solving,Crosson2014},
exactly solvable systems \cite{somma2012quantum,Roland2002}, and specific quantum chemistry
or state preparation problems
\cite{aspuru2005simulated,yung2014transistor,mcclean2013feynman,peruzzo2013variational}.
In contrast, recent results in the context of experimental quantum annealing machines,
which operate according to the same principle of adiabatic QC but in a thermal
environment, showed no evidence of quantum speedup for random optimization problems
\cite{Ronnow2014}.

To shed light on the actual power of QC, it is of great importance to be able to
perform extensive numerical and theoretical studies on large quantum systems.
Unfortunately, these kinds of analyses are strongly limited by the exponential
growth in dimensionality of quantum systems. In the context of adiabatic QC,
the evolution time, and thus the computational effort it quantifies, is related
to the minimum energy gap between the ground state and the first excited state
along the quantum evolution.
The direct calculation of the energy gap is feasible only for optimization
problems up to $n\approx 30$ qubits \cite{Farhi2001a,Hogg2003,Perdomo-Ortiz2010,Crosson2014}.
Estimations through quantum Monte Carlo techniques work only at finite temperature
and require a large overhead due to the equilibration and evolution steps necessary
to describe the situation at every stage of the adiabatic process
\cite{santoro2002theory,battaglia2006optimization,Young2008,hen2012excitation}.

\begin{table*}[t!]
\label{tab:summary}
\begin{tabular}{ccc}
\textbf{Driver Hamiltonian} & \textbf{Problem Hamiltonian} & \textbf{Dimensionality}\\[0.1cm]
\hline
\hline\\[-0.1cm]
& Grover Problem & \\[0.1cm]
  \hfsetfillcolor{red!10}
	\hfsetbordercolor{red!30}
	$\tikzmarkin{a}(0.2,-0.15)(-0.2,0.40)H_D^{(G)}\tikzmarkend{a}$
&
	\hfsetfillcolor{blue!10}
	\hfsetbordercolor{blue!30}
	$\tikzmarkin{aa}(0.2,-0.15)(-0.2,0.40)-\KetBra{\sigma^*}{\sigma^*}\tikzmarkend{aa}$
&
  2\\[0.3cm]
& Grover Problem & \\[0.1cm]
  \hfsetfillcolor{red!10}
	\hfsetbordercolor{red!30}
	$\tikzmarkin{b}(0.2,-0.15)(-0.2,0.40)H_D^{(S)}\tikzmarkend{b}$
&
	\hfsetfillcolor{blue!10}
	\hfsetbordercolor{blue!30}
	$\tikzmarkin{bb}(0.2,-0.15)(-0.2,0.40)-\KetBra{\sigma^*}{\sigma^*}\tikzmarkend{bb}$
&
  n+1\\[0.3cm]
& Grover Problem with Many Solutions (see Appendix~\ref{sec:multi_sol_Grover}) & \\[0.1cm]
  \hfsetfillcolor{red!10}
	\hfsetbordercolor{red!30}
	$\tikzmarkin{bz}(0.2,-0.15)(-0.2,0.40)H_D^{(S)}\tikzmarkend{bz}$
&
	\hfsetfillcolor{blue!10}
	\hfsetbordercolor{blue!30}
	$\tikzmarkin{bbz}(0.2,-0.15)(-0.2,0.40)-\sum_{i=1}^p\KetBra{\sigma^*_i}{\sigma^*_i}\tikzmarkend{bbz}$
&
  $\leq$ p$\,$n\\[0.3cm]
& Grover Problem with local noise & \\[0.1cm]
  \hfsetfillcolor{blue!10}
	\hfsetbordercolor{blue!30}
	$\tikzmarkin{f}(0.2,-0.15)(-0.2,0.40)H_D^{(G)}\tikzmarkend{f}$
&
	\hfsetfillcolor{blue!10}
	\hfsetbordercolor{blue!30}
	$\tikzmarkin{cc}-\KetBra{\sigma^*}{\sigma^*}\tikzmarkend{cc}$
	$\ \,+$
	\hfsetfillcolor{red!10}
	\hfsetbordercolor{red!30}
	$\tikzmarkin{dd}\sum_{i=1}^n\epsilon_i\hat\sigma_i^z\tikzmarkend{dd}$
&
  n+2\\[0.3cm]
& Grover Problem with local noise & \\[0.1cm]
  \hfsetfillcolor{red!10}
	\hfsetbordercolor{red!30}
	$\tikzmarkin{c}(0.2,-0.15)(-0.2,0.40)H_D^{(S)}\tikzmarkend{c}$
&
	\hfsetfillcolor{blue!10}
	\hfsetbordercolor{blue!30}
	$\tikzmarkin{ee}-\KetBra{\sigma^*}{\sigma^*}\tikzmarkend{ee}$
	$\ \,+$
	\hfsetfillcolor{red!10}
	\hfsetbordercolor{red!30}
	$\tikzmarkin{ff}\sum_{i=1}^n\epsilon_i\hat\sigma_i^z\tikzmarkend{ff}$
& n+1\\[0.3cm]
& Arbitrary M-level energy problems (including Random Energy Model) & \\[0.1cm]
  \hfsetfillcolor{blue!10}
	\hfsetbordercolor{blue!30}
	$\tikzmarkin{g}(0.2,-0.15)(-0.2,0.40)H_D^{(G)}\tikzmarkend{g}$
&
	\hfsetfillcolor{red!10}
	\hfsetbordercolor{red!30}
	$\tikzmarkin{gg}\sum_{i=1}^M E_i \sum_{\sigma\in\omega_{E_i}}\KetBra{\sigma}{\sigma}\tikzmarkend{gg}$
& M \\[0.3cm]
& Tunneling model with random barriers (see Appendix~\ref{sec:app_tunn_toy_model}) & \\[0.1cm]
  \hfsetfillcolor{red!10}
	\hfsetbordercolor{red!30}
	$\tikzmarkin{d}(0.2,-0.15)(-0.2,0.40)H_D^{(S)}\tikzmarkend{d}$
&
	\hfsetfillcolor{red!10}
	\hfsetbordercolor{red!30}
	$\tikzmarkin{hh}-\sum_{i=1}^n\hat\sigma_i^z\tikzmarkend{hh}$
	$\ \,+$
	\hfsetfillcolor{blue!10}
	\hfsetbordercolor{blue!30}
	$\tikzmarkin{ii}\sum_{\sigma|\omega(\sigma)=1}V_\sigma\KetBra{\sigma}{\sigma}\tikzmarkend{ii}$
&
  $\leq$(n+2)\textsuperscript{2}\\[0.3cm]
& Tunneling model with random barriers and local noise & \\[0.1cm]
  \hfsetfillcolor{red!10}
	\hfsetbordercolor{red!30}
	$\tikzmarkin{e}(0.2,-0.15)(-0.2,0.40)H_D^{(S)}\tikzmarkend{e}$
&
	\hfsetfillcolor{red!10}
	\hfsetbordercolor{red!30}
	$\tikzmarkin{jj}-\sum_{i=1}^n\hat\sigma_i^z\tikzmarkend{jj}$
	$\ \,+$
	\hfsetfillcolor{blue!10}
	\hfsetbordercolor{blue!30}
	$\tikzmarkin{kk}\sum_{\sigma|\omega(\sigma)=1}V_\sigma\KetBra{\sigma}{\sigma}\tikzmarkend{kk}$
	$\ \,+$
\hfsetfillcolor{red!10}
	\hfsetbordercolor{red!30}
	$\tikzmarkin{ll}\sum_{i=1}^n\epsilon_i\hat\sigma_i^z\tikzmarkend{ll}$
&
  $\leq$(n+1)\textsuperscript{3}\\[0.3cm]
\hline
\end{tabular}
\caption{%
\textbf{Examples where the proposed method gives an exponential reduction. Light-shaded boxes (red on-line)
and dark-shaded boxes (blue on-line) correspond
to {\boldmath $H_A$} and {\boldmath $H_B$} respectively as explained in the main text.} The first
column indicates the choice of the driver Hamiltonian corresponding to either
the Grover-style {\small $H_D^{(G)} = -\KetBra{\psi_0}{\psi_0}$} or the standard one
{\small $H_D^{(S)} = -\sum_{i=1}^n \hat{\sigma}_i^x$}. The second column describes
the optimization problem and
the third column provides an upper bound on the dimensionality after the
reduction method for a system of {$n$} qubits (to be compared with the total number
of state {$N=2^n$}).
The explanation of the symbols is as follows:
{$\Ket{\sigma}$} is the state of the computational basis corresponding to the
{$n$}-bit string {$\sigma\in\{0,1\}^n$}, {$\Ket{\psi_0}$} is the balanced superposition
of all the computational basis states, {$\pi(\cdot)$} is a permutation of {$\{1,2,\ldots,n\}$},
{$w(\cdot)$} the Hamming weight of a bit string, {$\hat\sigma_i^z$} and {$\hat\sigma_i^x$}
are respectively the Pauli $X$ and Pauli $Y$ matrices acting on the {$i-$}th qubit,
{$\Omega_E$} is the eigenspace associated with eigenvalue {$E$}, {$\epsilon_i = \pm |\epsilon|$} and
{$|\epsilon|,\,E,\,V_\sigma$} are real coefficients.
}
\end{table*}

In the past few years, several studies introduced special-purpose techniques to
reduce the dimensionality of particular classes of problems that are based on
explicit symmetries of the QC Hamiltonian. For example, algorithms
involving the Grover-style driver Hamiltonian have been analyzed in the subspace of
states symmetric under the exchange of any two qubits \cite{Roland2002,Znidaric2006,Hen2014a},
while cost functions that depend only on the Hamming weight of $n$-bit strings
have been solved by reducing the system to an effective single
spin $n/2$ \cite{Farhi2002b}. However, no clear
way to extend such approaches to non-symmetric situations has been suggested.
Here, we propose and implement a novel method to study large adiabatic
quantum optimizers by reducing the dimensionality of their Hilbert spaces. Our approach does
not rely on any explicit symmetry and goes beyond the strict distinction of driver
and problem contributions to the Hamiltonian (see Table~1).

The development of the present method allows us to perform the exact calculation of
the minimum gap for systems outside the usual assumption of an ideal, isolated adiabatic
quantum optimizer. In this direction, only few studies on simplified 2-level
systems have addressed the effect of thermal noise on adiabatic quantum optimization (AQO)
\cite{amin2008thermally}.
Here, we apply the dimensionality reduction to the Grover search problem in presence of
stochastic local noise (see Table~1), using two common choices of the driver Hamiltonian.
We are able to show that a quantum speedup is retained when an appropriate schedule,
independent of the choice of the target state, is implemented. To our knowledge, these
are the only conclusive results on the performance of adiabatic QC in presence of local
noise that has been reported so far, together with works on the effect of thermal baths
\cite{amin2008thermally,Amin2009} and on the specific D-wave hardware \cite{Dickson2013,Pudenz2014}.\\

The rest of the article is structured as follows: In Section~\ref{sec:AQO}, we introduce
the adiabatic quantum optimization and the main relevant quantities. In Section~\ref{sec:general-case} and
Section~\ref{sec:derivation}, we present our method and provide its detailed derivation.
The application of our method to the noisy Grover problem is then described in Section~\ref{sec:results},
while in the last Section we provide final discussions and conclusions.

\section{Adiabatic quantum optimization}
\label{sec:AQO}

In adiabatic quantum optimization, computational problems can be rephrased in terms of
finding those states which minimize a classical cost function encoded by a diagonal Hamiltonian.
The adiabatic theorem \cite{Messiah1958,Jansen2007} implies that a quantum system remains
in its instantaneous ground state if the quantum Hamiltonian is slowly deformed.
Following the above considerations, Farhi \emph{et al.} \cite{Farhi2001a} proposed to
govern the dynamics of a quantum optimizer by a time dependent Hamiltonian of the form:
\begin{equation}\label{eq:HamTotal}
    H_{\text{AQO}}(s) = (1-s(t))\, H_D \,+\, s(t)\, H_P,
\end{equation}
with $H_D$ being the initial Hamiltonian (usually called driver) and $H_P$ the Hamiltonian
associated to the problem to be optimized. The interpolation between the two Hamiltonians
takes a total time $T$
and is characterized by the adiabatic schedule $s(t)$ satisfying the boundary conditions
$s(0)=0$ and $s(T)=1$. With the system initially in the ground state of $H_D$, supposed
to be known and easy to prepare, the schedule will slowly drive it
to the ground state of $H_P$ at $t = T$. The question is how slowly the Hamiltonian
$H_{\text{AQO}}(s)$ must change to satisfy the adiabatic condition.

For problems that can be expressed as cost functions on $n$-bit strings, the problem
Hamiltonian is of the form
\begin{equation}\label{eq:ham_P}
    H_P = \sum_{\sigma\in\{0,1\}^n} E_\sigma \KetBra{\sigma}{\sigma},
\end{equation}
where $E_\sigma$ is the classical cost function of the configuration
$\sigma=\{\sigma_1,\sigma_2,\ldots,\sigma_n\}$ with \mbox{$\sigma_i\in\{0,1\}$}.
$E_\sigma$ represents the energy, according to the Hamiltonian $H_P$, of the quantum
state $\Ket{\sigma}$ expressed in the computational basis.
The solution of the optimization problem is provided by those states $\Ket{\sigma^\prime}$
associated to the lowest energy $E_{\sigma^\prime}\leq E_\sigma, \forall \sigma$.

The driver Hamiltonian $H_D$ can assume a variety of forms, but only a few regularly
appear in the literature: The ``Grover-style'' driver Hamiltonian (or simply
Grover driver Hamiltonian),
\begin{equation}\label{eq:ham_Dg}
    H_D^{(G)} = -\KetBra{\psi_0}{\psi_0},
\end{equation}
with $\Ket{\psi_0}=\frac{1}{\sqrt{2^n}} \sum_\sigma \Ket{\sigma}$ corresponding to
the equal superposition of all the states $\Ket{\sigma}$, and the ``standard'' driver
(corresponding to a transverse field)
\begin{equation}\label{eq:ham_Ds}
    H_D^{(S)} = -\sum_{i=1}^n \hat{\sigma}_i^x,
\end{equation}
where $\hat{\sigma}_i^x$ is the $X$ Pauli matrix acting on the $i$-th qubit, which physically
corresponds to a quantum transverse field. Despite their diversity, both $H_D$ are invariant
under the exchange of any pair of qubits, have the same ground state $\Ket{\psi_0}$
and do not commute with any $H_P$ apart from the trivial $H_P\propto\mathds{1}$ case.

The computational cost of AQO is quantified by the time one has to wait to obtain the
answer from the optimizer. If a single optimization run is performed, the computational
time $T_\text{comp}$ corresponds to the evolution time $T$ necessary to satisfy the
adiabatic condition to the desired precision.
In particular, a widely adopted condition \cite{Messiah1958} implies
$T\propto 1/{g_{\text{min}}^2}$, with $g_{\text{min}} = \min_s g(s)$ being the minimum
spectral gap between the ground state energy and the first excited state energy of
the adiabatic quantum Hamiltonian $H_{\text{AQO}}(s)$.
We calculate the computational time $T_\text{comp}$ in a more general way, discussed
in detail in Section~\ref{sec:Tcomp}, that takes into account the possibility of
performing multiple optimization runs with a shorter evolution time \cite{Boixo2014}.

\section{Dimensionality reduction method}
\label{sec:general-case}

The present method draws inspiration from the work of Roland and Cerf
\cite{Roland2002} in which the authors were able to obtain the exact spectral gap
for the Grover search problem on an adiabatic quantum computer by reducing the analysis
to an effective two-level system. We extend their approach in several directions,
to include arbitrary problem Hamiltonians, different choices of the driver Hamiltonian
and to deal with situations that do not present any explicit symmetry.

As a first step to reduce the effective dimensionality of the
Hilbert space, we rearrange the total Hamiltonian in Eq.~\eqref{eq:HamTotal}
in two distinct contributions
\begin{align}\label{eq:HamTotal2}
    H_{\text{AQO}}(s) &= (1-s(t))\, H_D \,+\, s(t)\, H_P \nonumber \\
                      &= a(s) \, H_A(s) \, + \, b(s) \, H_B(s) \, ,
\end{align}
where $H_A(s)$ and $H_B(s)$ do not necessarily correspond to the initial driver
or problem Hamiltonian and, in general, depend non-linearly on $s$.
To keep the notation as readable as possible, we will
omit any further dependence on $s$ when it is clear from the context.

Among the many possible choices of $H_A$ and $H_B$, the main idea is to
search for those combinations such that $H_A$ is a highly degenerate Hamiltonian
(with only $M$ distinct energy levels) and $H_B$ is a sum of $k$ rank-1 projectors, namely
\begin{subequations}
\begin{align}
\label{eq:ham_AB_A}
    &H_A = \sum_{E=1}^M E \, P_{\Omega_E} \, , \\
\label{eq:ham_AB_B}
	&H_B = \sum_{\alpha=1}^k \chi_\alpha \KetBra{\psi_\alpha}{\psi_\alpha},
\end{align}
\end{subequations}
with $\chi_\alpha\neq 0$ and $\{\Ket{\psi_\alpha}\}_{\alpha=1,\,\ldots,\,k}$
orthonormal states.
$\Omega_E$ is the subspace associated with the eigenvalue
$E$ of $H_A$ and $P_{\Omega_E}$ the corresponding projector.
The proposed method will lead to an exponential reduction of the effective dimension
of the Hilbert space whenever both $k$ and $M$ depend polynomially on the number of
qubits $n$. It is important to stress that the two Hamiltonians $H_A$ and $H_B$
do not necessarily commute and, therefore, their linear combination cannot be trivially
expressed as the sum of a polynomial number of orthogonal projectors.
At the moment, no automatic procedure exists to identify the most appropriate division
of $H_{\text{AQO}}$ and, therefore, one has to proceed by direct inspection.
Several examples are provided in Table~1.
\\


In the next Section, we show that the Hamiltonian {\small $H_{\text{AQO}}(s)$} has
a hidden block diagonal structure that appears evident when the basis is
chosen to include the states $\Ket{E_\alpha} \propto P_{\Omega_E}\Ket{\psi_\alpha}$.
Restricting the action of the Hamiltonian to the only block of dimension
larger than one, we obtain
\begin{align}\label{eq:ham_AQO2}
	H_{\text{eff}}(s) &=  a(s) \sum_E \sum_{\mu=1}^{\kappa(E)} E
			\KetBra{\mathcal{E}^{(E)}_\mu}{\mathcal{E}^{(E)}_\mu}\nonumber\\
		&\, \,+ b(s)\sum_{E,E^\prime}\sum_{\alpha=1}^k
		\chi_\alpha\mathcal{Z}_\alpha(E)\mathcal{Z}_\alpha(E^\prime)
			\KetBra{E_\alpha}{E^\prime_\alpha} ,
\end{align}
where $\mathcal{Z}_\alpha(E) = \|P_{\Omega_E}\Ket{\psi_\alpha}\!\|$ is a normalization
factor and the states $\{ \Ket{\mathcal{E}^{(E)}_\mu}\}_{\mu=1,\ldots,\kappa(E)}$ are given
by the orthogonalization of the set $\{ \Ket{E_\alpha} \}_{\alpha=1,\ldots,k}$. Here,
$\kappa(E)\leq k$ is the actual number of linearly independent $\Ket{\mathcal{E}^{(E)}_\mu}$
at given energy $E$.
For the sake of simplicity, in the following we will not explicitly indicate the
dependence on $E$ for the states $\Ket{\mathcal{E}_\mu}$.
As a consequence, the Hamiltonian in Eq.~\eqref{eq:ham_AQO2}
results to be an effective $(K \times M)$-level Hamiltonian, where
$K = \frac{1}{M}\sum_{E}\kappa(E)\leq k$.
We want to emphasize that the effective Hamiltonian is not an approximated
version of the original $H_{\text{AQO}}(s)$, but an exact description of its relevant
part.
In fact, if we extend the set $\{ \Ket{\mathcal{E}_\mu} \}_{E,\mu}$ to a
complete basis by adding orthonormal vectors belonging to eigensubspaces of $H_A$,
then $H_{\text{AQO}}(s)$ presents a block diagonal structure when
represented in such basis: The only block with dimension larger than
$1\times 1$ is a $(K\,M)\times (K\,M)$ block exactly reproduced by $H_{\text{eff}}$.

\section{Derivation of the effective Hamiltonian}
\label{sec:derivation}

In the previous Section, we started our analysis with the decomposition of the total
Hamiltonian for adiabatic quantum optimization (AQO) as the sum of two contributions,
$H_A$ and $H_B$, and expressed them in the form given by Eq.~\eqref{eq:ham_AB_A} and
Eq.~\eqref{eq:ham_AB_B}. Inserting such expressions in Eq.~{\eqref{eq:HamTotal2}} gives:
\begin{align}\label{eq:app_ham_AQO}
    H_{\text{AQO}}(s) &= a(s) \, H_A(s) \, + \, b(s) \, H_B(s) \nonumber \\
        &= a(s) \, \left[ \sum_{E=1}^M E \, P_{\Omega_E} \right]\, + \,
          b(s) \, \left[ \sum_{\alpha=1}^k \chi_\alpha \KetBra{\psi_\alpha}{\psi_\alpha}
                    \right] \, ,
\end{align}
where $E$ represents one of the $M$ distinct eigenvalues and
$P_{\Omega_E}$ the associated eigensubspace whose degeneracy is denoted by $\lambda(E)$.
We are seeking for a highly degenerate Hamiltonian $H_A$, with only $M$ distinct
energies, and an Hamiltonian $H_B$ formed by a small number $k$ of rank-1 projectors.
Here, we provide the justification of the claim that the relevant part of the energy
spectrum of $H_{\text{AQO}}(s)$ could be obtained studying an effective $M\times k$ system.
Initially, we present the derivation in the case in which $k=1$, \emph{i.e.} for a
Grover-style Hamiltonian $H_B=-\KetBra{\psi_1}{\psi_1}$, since the procedure is more
intuitive.

\subsection{Special case $\boldsymbol{k=1}$}
\label{sec:M1_ham}

Consider the case in which $H_B$ corresponds to a single rank-one projector.
The extension to the general case is presented after the restricted case $k=1$.
From the completeness of $H_A$ we have $\sum_E P_{\Omega_E}=\mathds{1}$ and
$\sum_E \lambda(E)=2^n$. For each energy $E$, we define
$\Ket{E}=\frac{P_{\Omega_E} \Ket{\psi_1}}{\mathcal{Z}(E)}$
as the normalized projection of $\ket{\psi_\alpha}$ on the subspace $P_{\Omega_E}$,
and introduce $[\lambda(E)-1]$ orthonormal states to obtain a basis of $\Omega_E$:
$\left\lbrace\Ket{E},\,\Ket{E_1^\perp},\,\ldots,\,\Ket{E_{\lambda(E)-1}^\perp}\right\rbrace$.
We have
\begin{equation}\label{eq:app_new_proj}
	P_{\Omega_E} = \KetBra{E}{E}+\sum_{i=1}^{\lambda(E)-1}\KetBra{E_i^\perp}{E_1^\perp}
\end{equation}
and then
\begin{subequations}\label{eq:app_ham_AQO2}
\begin{align}
	H_{\text{AQO}} &= -b(s)\KetBra{\psi_1}{\psi_1}\label{eq:app_ham_AQO2_1}\\
		&\ \ \ + a(s) \sum_E^{\phantom{1}} E \KetBra{E}{E}\label{eq:app_ham_AQO2_2}\\
		&\ \ \ + a(s) \sum_E E \sum_{i=1}^{\lambda(E)-1}\KetBra{E_i^\perp}{E_i^\perp}\,.\label{eq:app_ham_AQO2_3}
\end{align}
\end{subequations}
Notice that, while $\Braket{\psi_1|E}$ can be non-zero, $\Ket{\psi_1}$ and $\Ket{E_i^\perp}$
are always orthogonal because
\begin{subequations}
\begin{align}\label{eq:app_orth_cond2}
	P_{\Omega_E}\Ket{E_i^\perp} &= \Ket{E_i^\perp},\\
	P_{\Omega_E}\Ket{\psi_1^{\phantom{\perp}}} &= \mathcal{Z}(E)\Ket{E},
\end{align}
\end{subequations}
and then
\begin{align}\label{eq:app_orth_cond3}
	\Braket{\psi_1 | E_i^\perp} &= \Braket{\psi_1 | P_{\Omega_E} | E_i^\perp } \nonumber\\
		&= \mathcal{Z}(E) \Braket{E | E_i^\perp} = 0 \,.
\end{align}

We observe that Eq.~\eqref{eq:app_ham_AQO2} describes an Hamiltonian that is block diagonal
in the basis
\begin{equation}
	\bigcup_E \left\{\Ket{E},\,\Ket{E_1^\perp},\,\ldots,\,\Ket{E_{\lambda(E)-1}^\perp}\right\},
\end{equation}
since the terms in Eq.~\eqref{eq:app_ham_AQO2_3} act on different subspaces with respect
to the terms in Eq.~\eqref{eq:app_ham_AQO2_1} and Eq.~\eqref{eq:app_ham_AQO2_2}.
Thus, the relevant part of the AQO Hamiltonian results
\begin{align}\label{eq:app_ham_AQO3}
	H_{\text{eff}} &= -b(s) \KetBra{\psi_1}{\psi_1} +
		              a(s) \sum_E^{\phantom{1}} E \KetBra{E}{E} \nonumber \\
                   &= -b(s) \sum_{E,E^\prime} \mathcal{Z}(E) \mathcal{Z}(E^\prime)
                      \KetBra{E}{E^\prime} +
		              a(s) \sum_E^{\phantom{1}} E \KetBra{E}{E} \, ,
\end{align}
which is an effective $M-$level Hamiltonian, where $M$ is the number of distinct energy
levels of the contribution $H_A$.

\subsection{General case}
\label{sec:Mk_ham}

Here, we present the derivation of our reduction method in the general case of
arbitrary $M$ and $k$. We will, then, show that it is always possible to reduce
a generic AQO Hamiltonian to a $(M \times K)-$level Hamiltonian, where $M$ is
the number of energies of $H_A$ and $K$ is an integer number equal or smaller
than the number $k$ of states over which the term $H_B$ acts non-trivially.
Let us consider the Hamiltonian in Eq.~\eqref{eq:ham_AB_B}.
With a straightforward generalization of the notation, we introduce
\begin{equation}
	\Ket{E_\alpha} = \frac{P_{\Omega_E}\Ket{\psi_\alpha}}{\mathcal{Z}_\alpha(E)},
\end{equation}
with $\mathcal{Z}_\alpha(E) = \|P_{\Omega_E}\Ket{\psi_\alpha}\|$
and divide the subset $\Omega_E$ in two parts, one spanned by
$\{\Ket{E_\alpha}\}_{\alpha=1\,\ldots,\,k}$ and the other representing its
orthogonal complement $\omega_E$. As for the $1-$state case, the set
$\omega_E$ is by construction contained in the kernel of $H_B$, such that
\begin{equation}
	\Braket{E^\perp|\psi_\alpha} = 0,
\end{equation}
for any $\Ket{E^\perp}\in\omega_E$ and for any energy $E$.
As a consequence, all states in $\omega_E$ can be neglected in the effective AQO Hamiltonian. Moreover, since it is not said that $\Braket{E_\alpha|E_\beta} = \delta_{\alpha\beta}$,
we use the orthogonalization procedure presented in the next Section to extract from the
original set $\{\Ket{E_\alpha}\}_{\alpha=1,\,\ldots,\,k}$ a smaller set of
$\kappa(E) \leq \min\{k,\,\lambda(E)\}$ orthonormal states
$\{\Ket{\mathcal{E}_\mu}\}_{\mu=1,\,\ldots,\,\kappa(E)}$.

In this way
\begin{equation}
	P_{\Omega_E} = \sum_{\mu=1}^{\kappa(E)} \KetBra{\mathcal{E}_\mu}{\mathcal{E}_\mu} +
		\sum_{\Ket{E^\perp}\in\omega_E}\KetBra{E^\perp}{E^\perp},
\end{equation}
and recalling that
\begin{align}
	\Ket{\psi_\alpha} &= \left(\sum_E P_{\Omega_E}\right)\Ket{\psi_\alpha} = \sum_E \mathcal{Z}_\alpha(E) \Ket{E_\alpha} ,
\end{align}
the (relevant part of the) AQO Hamiltonian in Eq.~\eqref{eq:app_ham_AQO} becomes
\begin{align}\label{eq:app_ham_AQO5}
	H_{\text{eff}} &= b(s) \sum_{\alpha=1}^k \chi_\alpha \sum_{E,E^\prime} \mathcal{Z}_\alpha(E)\mathcal{Z}_\alpha(E^\prime)
		\KetBra{E_\alpha}{E^\prime_\alpha} \nonumber \\
		& \ \ + a(s) \sum_E E \sum_{\mu=1}^{\kappa(E)} \KetBra{\mathcal{E}_\mu}{\mathcal{E}_\mu}.
\end{align}
In the equation above, we already removed all terms in $\omega_E$ because
they are factorized with respect to the relevant part of the AQO Hamiltonian.
As one can see, Eq.~\eqref{eq:app_ham_AQO5} describes an effective
$(M\times K)$-level Hamiltonian, where $K = \frac{1}{M}\sum_{E}\kappa(E)$.
Correctly, if $k = 1$ we obtain the AQO Hamiltonian reported in Eq.~\eqref{eq:app_ham_AQO3}.

It is important to observe that we reduced the original AQO Hamiltonian in
Eq.~\eqref{eq:app_ham_AQO} to an effective $(M\times K)-$level Hamiltonian, and
then we reduced the Hilbert space from $2^n$ states to $(M\times K)$ states.
Therefore, if both $K$ and $M$ are polynomial in the number of spins $n$, the
reduced AQO Hamiltonian in Eq.~\eqref{eq:app_ham_AQO5} can be expressed using only a polynomial number of states, that is
to say that we obtained an exponential reduction of the Hilbert space. We observe that
the calculation of $\mathcal{Z}_\alpha(E)$ and $\Ket{E_\alpha}$ might be non trivial for
arbitrary states $\Ket{\psi_\alpha}$ and Hamiltonian $H_A$.

\subsection{Orthogonalization procedure of $\boldsymbol{\{\Ket{E_\alpha}\}}$}
\label{sec:orthogonalization}

The states $\Ket{E_\alpha}$ are, in general, not orthogonal but they can
be expanded as a linear combination of the orthonormal
states $\{\Ket{\mathcal{E}_\mu}\}$ which, we recall, span the effective subspace
containing the relevant part of the total energy spectrum. Here, we present the
mathematical procedure to perform the orthogonalization.
Introducing the $\kappa(E)\times k$ matrix $T$ with entries
$T_{\mu \alpha}=\Braket{\mathcal{E}_\mu|E_\alpha}$, one has:
\begin{equation}
	\Ket{E_\alpha} = \sum_\mu \Braket{\mathcal{E}_\mu|E_\alpha} \Ket{\mathcal{E}_\mu}
			     = \sum_\mu T_{\mu \alpha} \Ket{\mathcal{E}_\mu} .
\end{equation}
Then, we can write:
\begin{align}
	\Braket{E_\alpha|E_\beta} &= \sum_{\mu ,\nu} \Braket{ \mathcal{E}_\mu | T_{\mu \alpha}^* T_{\nu \beta} | \mathcal{E}_\mu } \nonumber \\
					      &= \sum_\mu T^\dagger_{\alpha \mu} T_{\mu \beta} = [T^\dagger T]_{\alpha \beta} ,
\end{align}
and interpret the above values as the entries of a certain matrix $V$.
Such matrix is a square matrix with linear dimension $k$ and can be shown to be Hermitian
and positive-semidefinite. Therefore it admits a Cholesky decomposition:
\begin{equation}
	V = U^\dagger U
\end{equation}
where $T$ is an upper triangular matrix with real and positive diagonal entries. While every
Hermitian positive-definite matrix has a unique Cholesky decomposition, this does not need
to be the case for Hermitian positive-semidefinite matrices and this reflects a certain
freedom in choosing the states $\{\Ket{\mathcal{E}_\mu}\}$.
It appears clear that the expansion coefficients $T_{\mu \alpha}$ are the entries of a
particular choice of such matrix $U=T$.

Expressing the Hamiltonian $H_B$ in the basis of the effective subspace, we have
\begin{align}\label{eq:app_explicit-Hd}
      \Braket{\mathcal{E}_\mu|H_B|\mathcal{E}^\prime_\nu} &=
                 \sum_{\alpha=1}^k \chi_\alpha \Braket{\mathcal{E}_\mu|\psi_\alpha} \Braket{\psi_\alpha|\mathcal{E}^\prime_\nu} \nonumber \\
             &= \sum_{\alpha=1}^k \chi_\alpha \mathcal{Z}_\alpha(E) \mathcal{Z}_\alpha(E^\prime) \Braket{\mathcal{E}_\mu|E_\alpha}
                    \Braket{E^\prime_\alpha|\mathcal{E}^\prime_\nu} \nonumber \\
             &= \chi_\alpha \mathcal{Z}_\alpha(E) \mathcal{Z}_\alpha(E^\prime) \left( + \sum_{\alpha=1}^n T^{(E)}_{\mu \alpha} T^{(E^\prime)*}_{\nu \alpha} \right) \nonumber \\
             &= \chi_\alpha \mathcal{Z}_\alpha(E) \mathcal{Z}_\alpha(E^\prime)
                \left[ T^{(E)}T^{(E^\prime)\dagger} \right]_{\mu \nu}
\end{align}
whereas the term $H_A$ becomes
\begin{equation}\label{eq:app_explicit-Hp}
      \Braket{\mathcal{E}_\mu|H_A|\mathcal{E}^\prime_\nu} = E \delta_{E E^\prime} \delta_{\mu \nu} \, .
\end{equation}
In this way, we have expressed all the necessary operators in the reduced basis.\\


It is important to appreciate a subtlety: In most cases, we do not know the
exact form of the states $\Ket{{E}_\alpha}$, for example because they are related
to the eigenstates of $H_A$. Then, how can we obtain the
explicit entries of $H_{\text{eff}}$ in Eq.~\eqref{eq:ham_AQO2} to perform
the numerical analysis? The answer is indirectly contained in the detailed
derivation above since we showed that all the entries of $H_\text{eff}$ can
be computed from the knowledge of the overlap matrix $\Braket{E_\alpha|E_\beta}$
at a given energy $E$. For many relevant cases, such
overlaps can be computed either analytically or
numerically by means of algorithms which require only a polynomial amount of (spatial)
classical resources. To give an example, when the effective Hamiltonian depends
only on the degeneracy of the spectrum of $H_A$, usually called the density of states,
this information can be estimated using entropic sampling techniques
\cite{Lee1993,Wang2001,Dickman2011}.
More generally, Table~1 lists a few situations where the proposed
method can be applied to exponentially reduce the effective dimensionality of
$H_\text{AQO}$: As one can see, the suggested method can successfully represent
problems in which neither the problem nor the driver
Hamiltonian are of Grover-style form. In Section~\ref{sec:results}, we provide an
explicit example to illustrate how the proposed method works in the context of adiabatic
quantum optimization in presence of local noise.


\section{Applications}
\label{sec:results}
\subsection{Grover search problem with discrete disorder}
\label{sec:Grover-with-disorder}

In 1996, Grover introduced a quantum algorithm to search for target entries in
unstructured databases, demonstrating that quantum computers achieve a quadratic
speedup with respect to the best possible classical algorithm \cite{Grover1997}.
This fundamental result was later extended to adiabatic QC finding that it is
possible to reproduce the quadratic speedup if one tailors the
adiabatic schedule in such a way that $H_{\text{AQO}}(s)$ varies very slowly
only in correspondence of the smallest gap \cite{Roland2002}.
Indeed, such quadratic speedup represents the maximum speedup achievable
with AQO for unstructured searches or Grover-style Hamiltonians for which
$T_{comp} \geq O(\sqrt{2^n})$
\cite{Farhi1998,farhi2008make,ioannou2008limitations,cao2012efficiency}.

Ideally, the energy landscape associated to unstructured databases should be perfectly
flat, but this is not the case in realistic situations in which, for example, imprecision
in local control fields can give rise to a local disorder term.
It is not unreasonable to suspect that the quantum speedup might be diminished or
even lost due to this noise contribution or due to effects similar to Anderson
localization \cite{ Anderson1978}. Here, we apply
the proposed reduction method to study the Grover search problem in the presence of increasing
amounts of local disorder, using both the Grover like driver Hamiltonian and the
standard (transverse field) Hamiltonian. Our results show that adiabatic QC still remains
faster than any classical algorithm.

First of all, we have to specify the noise model. Several and diverse models have been
introduced in previous works related to the Grover search or AQO
\cite{Shenvi2003,Roland2005,Tiersch2006}: Here, we consider a local term of the form
\begin{equation}\label{eq:ham_loc_noise}
    H_{{dis}} = \sum_i \epsilon_i \hat{\sigma}_i^z \, ,
\end{equation}
in addition to the Grover-style problem Hamiltonian $-n\KetBra{\sigma^*}{\sigma^*}$, with
$\Ket{\sigma^*}$ being the target state (see Table~1). Observe that we rescale the energy
of the target state in order to keep it extensive with the system size.
For simplicity, we choose $\epsilon_i=\pm\epsilon$ with $\epsilon\geq0$ and the sign
randomly drawn with $50:50$ probability.
%
%
Notice that one obtains an exponential reduction in the dimensionality of the problem
even when the $\epsilon_i$ are allowed to assumes a finite set of distinct values
(see Appendix \ref{sec:noisy_hamiltonian}).
Even if the discrete noise model in Eq.~\eqref{eq:ham_loc_noise}
is simplistic, it qualitatively catches many of the results of a localized noise.

For the calculation, we assume that the disorder is static during a single adiabatic run,
but that it can vary between successive repetitions of the adiabatic algorithm
\cite{venturelli2014,king2014algorithm}. In the quantification of the computational
time associated to the quantum algorithm, we take into account the possibility
of repeating the run instead of increasing the single evolution time, see
Section~\ref{sec:Tcomp}. This approach is becoming standard in the adiabatic
QC literature \cite{Boixo2014,Ronnow2014}.

Second, to bring the Hamiltonian in the most suitable form, we apply local
$\hat{\sigma}_i^x$ operators to change the sign of the positive $\epsilon_i$.
The action of $U_x=\prod_{i \text{ s.t. } \epsilon_i<0} \hat{\sigma}_i^x$ leaves the
overall spectrum unchanged. Finally, we divide
the rotated total Hamiltonian in two parts (see Table~1):
\begin{align}\label{eq:Grover_disorder}
	U_x H_{\text{AQO}} U_x^\dagger
      &= U_x \left[ (1-s) \, H_D^{(S)}\,+\,s \, H_P \right] U_x^\dagger \nonumber \\
      &= -(1-s)\sum_{i=1}^n \hat{\sigma}_i^x \,
		 - \, s \sum_{i=1}^n |\epsilon_i| \hat{\sigma}_i^z \,
         - \, s \, n \, \KetBra{\sigma^\prime}{\sigma^\prime} \nonumber \\
	  &= -\gamma(s) {\sum_{i=1}^n \hat{{\sigma}}_i(s)}
            \, -s \, n \, {\KetBra{{\sigma^\prime}}{{\sigma^\prime}}} \nonumber \\
	  &= \gamma(s) \, {H_A} \,+\, s \, {H_B} \, ,
\end{align}
where $\Ket{{\sigma}^\prime} = U_x\Ket{\sigma^*}$ is the target of the rotated AQO
Hamiltonian, $\gamma(s)=\sqrt{(s\,\epsilon)^2 + (1-s)^2}$,
and every $\hat{{\sigma}}_i(s)=\tfrac{1-s}{\gamma(s)} \hat\sigma_i^x +
\tfrac{s\,\epsilon}{\gamma(s)} \hat\sigma_i^z$
is an identical single qubit operator which acts on the $i$-th qubit as a rotated
Pauli matrix. A derivation of the reduced Hamiltonian for
the Grover-style driver Hamiltonian is included in Appendix \ref{sec:noisy_hamiltonian2}.
We observe that more general situations, in which the direction of the noise
varies freely (and in a continuous way) for each distinct spin, can be included
by following an analogous approach. The only difference from
Eq.~\eqref{eq:Grover_disorder} is that the spin matrices $\hat\sigma_i(s)$ are
now rotated in distinct directions.

The drastic dimensionality reduction, from $2^n$ states to only $(n+1)$ states, allows
us to calculate the energy gap $g(s)$ at any point during the evolution. The results
are expressed in terms of the computational time $T_\text{comp}$, namely
the temporal cost for the quantum algorithm to reach the success probability of $99\%$
\cite{Boixo2014,Ronnow2014}.

\begin{figure}[t!]
  \centerline{\includegraphics[width=0.45\textwidth]{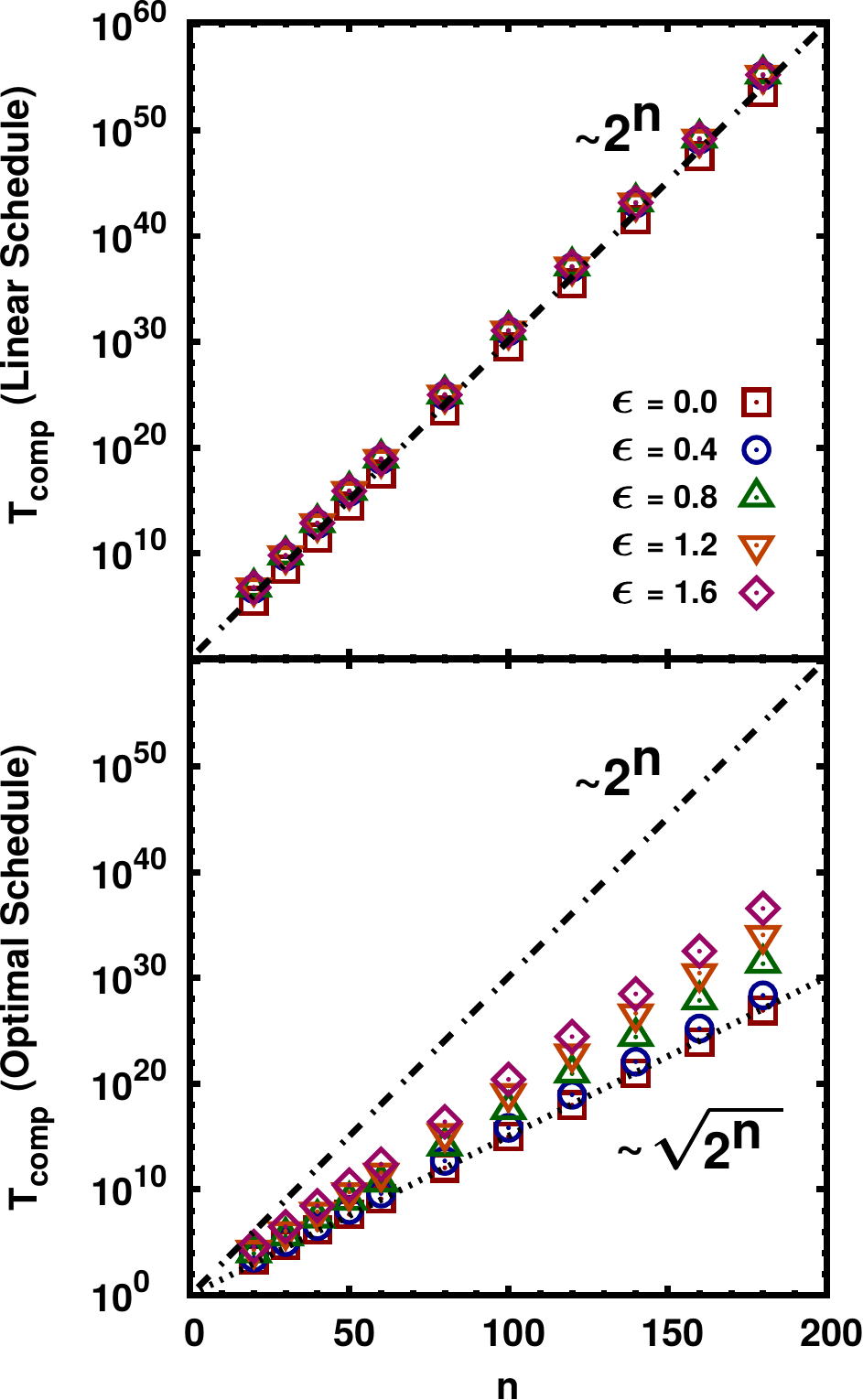}}
  \label{fig:disorderVSn}
  \caption{
    \textbf{Even in the presence of discrete local noise, the adiabatic QC is faster
    than any classical algorithm for searching an unstructured database.}
    The proposed method is applied to calculate the computational time necessary
    to solve the Grover search problem in presence of local disorder.
    The computational scaling is compared, for increasing strength of the local noise,
    to the best classical result ($T_\text{comp} \propto N$,
    where $N = 2^n$ is the number of entries in the database) and the best quantum
    result in the ideal case where the noise is absent ($T_\text{comp}~\propto \sqrt{N}$).
    We consider two annealing schedules, the linear one (Top panel) and an optimal schedule
    determined by imposing the adiabatic condition locally (Bottom panel). A quantum speedup
    is possible only when optimal schedules are adopted.
    These results are obtained by using the standard driver Hamiltonian, but similar
    curves have been also obtained for the Grover-style driver Hamiltonian.
    }
\end{figure}

\subsection{Calculation of the computational time}
\label{sec:Tcomp_Grover-with-disorder}

In general terms, the performance of an adiabatic quantum optimizer is expected to improve
if the evolution time is increased, since the conditions behind the adiabatic quantum
theorem are better satisfied. However, it may be possible that a larger probability of
success is achieved if the adiabatic quantum optimizer is used for a shorter evolution
time, but in repeated runs \cite{Boixo2014,Ronnow2014}. In Appendix ~\ref{sec:Tcomp}, we
provide a precise analysis of the computational time required to achieve
a solution in the general case of an arbitrary adiabatic quantum optimization.
To make the definition of computational time (see Appendix~\ref{sec:Tcomp}) more concrete,
let us apply it to the Grover problem with local disorder and calculate the computational
time according to Eq.~\eqref{eq:Tcomp2}.
Given a target state and a specific realization of the local disorder, the noise term
can either increase or decrease the target state energy according to the number $q\in[0,n]$
of spins where the disorder provides a positive energy contributions.
Assuming that $\epsilon_i = \pm\epsilon$ are randomly drawn with 50:50 probability, the
probability distribution for $q$ results
\begin{equation}
	p_n(q) = 2^{-n}\binom{n}{q}\,,
\end{equation}
where $n$ is the number of qubits, while the success probability reads
\begin{equation}
	p_S(q,\,T|\,q^*) = \delta(q-q^*)\,\Theta(T - T_\text{ann}(q^*))\,\Theta(q_\epsilon - q)\,,
\end{equation}
in which the second $\Theta$ function takes into account that the target state is the
ground state of the noisy Hamiltonian only for $q\leq q_\epsilon$, with
$q_\epsilon = \lfloor\frac{n}{2\epsilon}\rfloor$.

Unlike the case of the standard driver Hamiltonian for which the calculation of $T_\text{ann}$
is not trivial, for the Grover-style driver Hamiltonian we can provide an accurate estimate
of $T_\text{comp}$, given that $T_\text{ann} = \sqrt{2^n}$ regardless the problem Hamiltonian
\cite{farhi2008make}.
Recalling that $\lim_{n\to0} p_n(q) = 0$ (observe that even the mode of the distribution
$p_n(q)$ scales like $\max_{q} p_n(q) \approx 1/\sqrt{n}$), the computational times becomes
\begin{align}\label{eq:Tcomp3}
	T_\text{comp}(\epsilon)
        & =     \sqrt{2^n}\min_{q^*} \Big \{\frac{\log(1-0.99)}{\log(1-p_n(q^*)\Theta(q_\epsilon - q^*)))} \Big\} \nonumber \\
        & \approx \log(1-0.99)\min_{0\leq q^*\leq q_\epsilon}\Big\{2^{3n/2 - n\,h(q^*/n)}\Big\}\,,
\end{align}
where we used the Stirling approximation $\log_2\binom{n}{q} \approx n\,h(q/n)$ in which
\begin{equation}
	h(x) = -x \log_2 x - (1-x) \log_2 (1-x)
\end{equation}
is the Shannon entropy. Finally, the computational scaling is
\begin{equation}
	s(\epsilon) = \lim_{n\to\infty}\left[\frac{1}{n}\log_2 T_\text{comp}(\epsilon)\right] =
	\begin{cases}
		\frac{1}{2} & \, \epsilon < \tilde\epsilon\\
		\frac{3}{2} - h\left(\frac{1}{2\epsilon}\right) & \,  \text{otherwise}
	\end{cases},
\end{equation}
where $\tilde{\epsilon} = 1$ is the noise threshold such that the minimum energy of
$H_\text{dis}$ becomes comparable with the energy of the target state. Interestingly,
it exists a noise threshold $\epsilon^\text{cl} \approx 4.54$ such that
$s(\epsilon) \geq 1$ for $\epsilon > \epsilon^\text{cl}$,
aka the AQO cannot perform better than classical computers in that regime. Indeed, for
large $\epsilon$, the probability of the target state to be the true ground state of
$H_\text{AQO}$ with noise becomes smaller. Therefore, minimizing $H_\text{AQO}$ becomes
less efficient than simply trying to find the target state by an exhaustive enumeration.
We note that such effect is somewhat artificial since the success probability for very
short evolution times tends to $2^{-n}$ and not to zero as we, conservatively, assumed.\\

Observe that a more elaborate annealing schedule can partially remove the necessity
of repeating runs. In fact, if the annealing schedule $s(t)$ is chosen to be the
solution of the following equation
\begin{equation}
  \frac{ds}{dt} = \epsilon \min_q g^2(s,\, q),
\end{equation}
then it is guaranteed that the quantum dynamics is adiabatic, regardless the hidden
parameter $q$. The evolution time for a single run is expected to increase only
linearly in $n$ as compared to the schedule considered above.
However, for sufficiently large strength of the noise, fluctuations can affect the
energy landscape of the problem Hamiltonian with the consequence that the global
ground state of the noisy problem Hamiltonian is not anymore the desired target state.
In these cases, even for a very slow quantum adiabatic evolution, the final state
will not correspond to the target state regardless the evolution time.
By quenching the noise and repeating the evolution run such problem is naturally
solved since more favorable noise realizations are possible.

\subsection{Scaling analysis}
\label{sec:scaling-analysis}

\begin{figure}[t!]
  \centerline{\includegraphics[width=0.45\textwidth]{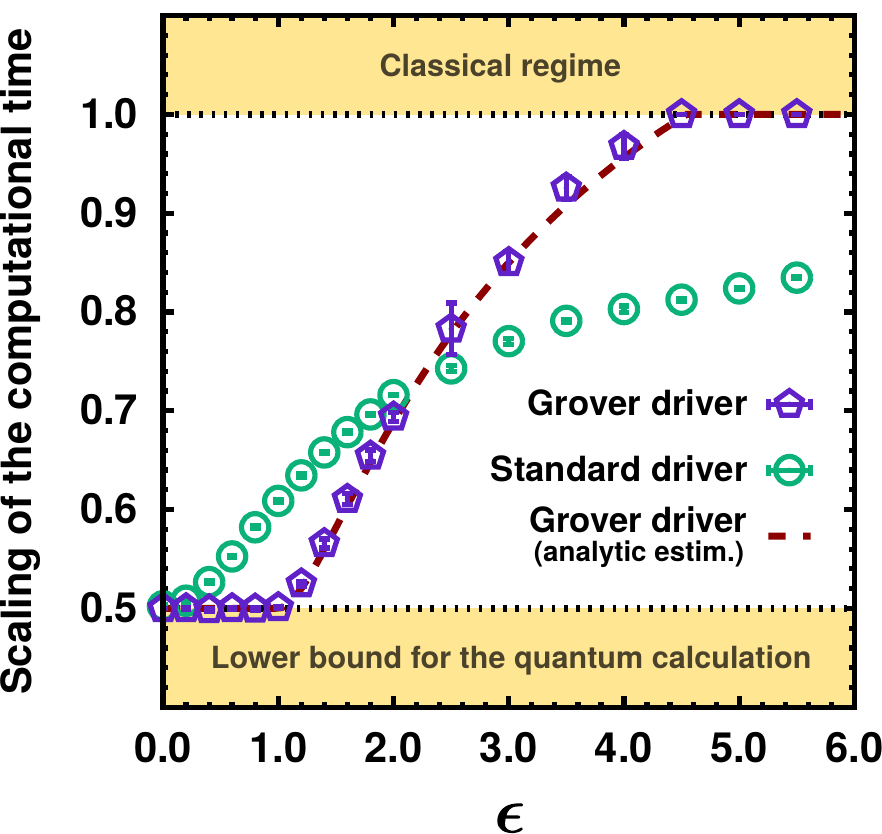}}
  \label{fig:disorderVSe}
  \caption{
    \textbf{Polynomial quantum speedups are retained even in the presence of local disorder.}
    The figure shows the behavior of the coefficient characterizing the exponential scaling for the Grover search
    problem against the strength of the disorder. We adopt an optimal schedule that
    would guarantee a quadratic speedup in absence of noise. We find that adiabatic QC still
    retains a better scaling than any classical algorithm even if the quantum speedup is
    reduced for increasing level of disorder. Interestingly, although the Grover-style
    driver Hamiltonian gives better performances for weak noise, the standard driver
    Hamiltonian results more ``robust'' for large noise. The exponential coefficient has been
    obtained by fitting $T_\text{comp}$ for systems up to $n \leq 160$ qubits.
    }
\end{figure}

In this Section we present our main results on the computational scaling of the
noisy Grover problem. Fig.~1 
shows the scaling behavior of $T_\text{comp}$ by varying the
level of noise, using either a linear schedule (Top) or an optimal schedule (Bottom)
tailored to the noise model, but independent of the specific target state $\Ket{\sigma^*}$
(see Appendix \ref{sec:noisy_hamiltonian} and \ref{sec:noisy_hamiltonian2}).
In both cases we employ the standard driver Hamiltonian.
We observe that, for the linear schedule, neither quantum speedup nor noise effects are
observed. The optimal schedule, instead, gives rise to a quadratic quantum speedup in the
noiseless case that is, interestingly, only partially canceled when local disorder is
taken into account.
We also compared the performance of an adiabatic quantum optimizer when the standard driver
Hamiltonian is substituted with the Grover-style driver Hamiltonian, in order to see how much
the choice of the driver influences the ``robustness'' of the AQO to local noise (see Fig.~2):
In this case, the Grover-style driver preserves all the quadratic quantum speedup if the
noise is maintained below a certain threshold $\epsilon \lesssim \tilde\epsilon$,
but the AQO speedup quickly degrades for moderate disorder until the classical scaling
is finally reached.
The turning point $\tilde\epsilon \approx 1$ corresponds to the noise threshold for which
the lowest energy of $H_\text{dis}$ is comparable with the energy of the target state
(see Appendix \ref{sec:noisy_hamiltonian} for more details).
Conversely, the standard driver Hamiltonian appears significantly more ``robust''
at large disorder, so that the performance of the adiabatic QC gently decreases for
increasing strength of the noise.
These are good news for the possibility of implementing adiabatic QC in realistic
systems since one can retain all the quantum speedup (for very weak disorder) or most
of it (for moderate disorder) by choosing the appropriate driver Hamiltonian.\\

In the following, we analyze the effect of different driver Hamiltonians by
observing the behavior of the minimum gap at various $q/n$ ratios for the specific
system size $n=160$. As a reminder, we have denoted by $q$ the number of spins
where the disorder provides a positive energy contributions. Fig.~3 
shows that the minimum gap is practically independent of both $q/n$ and
$\epsilon$ when the Grover driver is used: This could be expected since, for its nature,
$H_D^{(G)}$ does not see any underlying structure of the problem energy landscape, not even
the noise contribution, and presents a minimum gap only influenced by the degeneracy of the
ground state (in our case, we have a unique ground state as long as $q\leq q_\epsilon$).
On the contrary, the energy landscape plays a role during the adiabatic evolution with
$H_D^{(S)}$ and this can be easily observed for the special case $q=0$.
In this case, Eq.~\eqref{eq:H_AQO_noise1} assumes the form
\begin{align}\label{eq:H_AQO_noise1_q0}
	H^\prime_\text{AQO} &= -s n \KetBra{0}{0}
		- \sum_{i=1}^n \Big[s|\epsilon_i| \hat\sigma_i^z + (1-s)\hat\sigma_i^x\Big],
\end{align}
and the target state $\KetBra{0}{0}$ is \emph{also} the ground state of
$- \sum_{i=1}^n s|\epsilon_i| \hat\sigma_i^z$. For small $\epsilon\ll 1$ one recovers the
case of the noiseless Grover problem, while for large $\epsilon \gg 1$ the situation is
analogous to the Hamming weight problem that presents a gap largely independent of $n$.
The fact that the absolute value of the minimum gap at small $q/n$ is always larger for
the standard driver (even for $\epsilon<1$ when the scaling of the computational time
is better with the Grover driver) can be understood observing that the size of the
minimum gap is only one of three factors that influence $T_{\text{comp}}$;
the other two being the shape of the minimum gap (especially its width in the adiabatic
coordinate $s$) and the probability that a certain $q/n$ is realized in practice.
%

%
\begin{figure}[t!]
  \centering
  \includegraphics[width=0.45\textwidth]{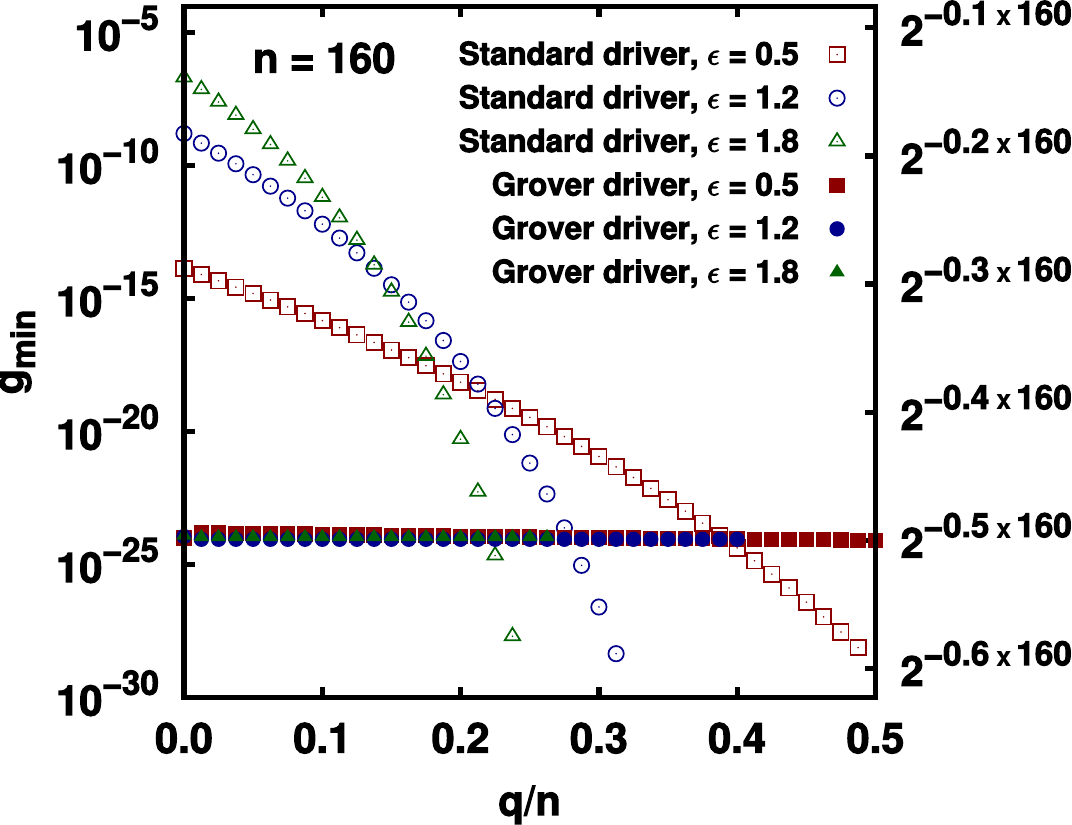}
  \caption{\label{fig:gap}
    Minimum gap calculated for both the standard driver and Grover driver Hamiltonian
    applied to the Grover search problem with local disorder.
    In the first case, $g_\text{min}$ spans several orders of magnitude if either $q$ or
    ${\epsilon}$ are varied. In the latter case, the minimum gap is, instead, almost
    constant and scales as ${g_\text{min} = 1/\sqrt{2^n}}$.
    This plot is obtained for systems of $n = 160$ qubits.
    The symbols are plotted only for those $q/n$ such that $q<q_\epsilon$.}
\end{figure}
%

\section{Conclusions}
\label{Conclusions}

Estimation of the computational power of the adiabatic quantum optimization
requires the knowledge of the spectral gap of the total Hamiltonian
$H_{\text{AQO}}(s)$. Its direct quantification is a hard task since it requires
the calculation of eigenvalues of matrices which are exponentially large in the
number of qubits. To circumvent this limitation, several methods have been
proposed to diminish the classical resources necessary to represent the adiabatic
quantum optimizer.
However, these special-purpose approaches are based on the exploitation of symmetries
of either the driver or the problem Hamiltonian, and are therefore confined
to particular classes of problems.

Here, we present and discuss a method that reduces the effective dimensionality
of the system even in absence of explicit symmetries, and that goes beyond the
idea of studying the properties and structure of the driver and problem terms
separately. Formally, this is made possible by the identification of a hidden
block diagonal structure in the total Hamiltonian and, consequently, by the
existence of a small subspace in which the relevant eigenstates are effectively
confined. According to the specific total Hamiltonian $H_\text{AQO}$,
the present method requires only the knowledge of quantities that can
be computed either analytically or by using efficient numerical approaches.

We apply the proposed method to calculate the energy gap, in
a numerically exact way, for large systems exposed to local disorder or other forms
of imprecision in the values of the parameter that characterize the problem
Hamiltonian: Interestingly, we show that adiabatic quantum computation seems
to be robust enough to deal with a form of stochastic local noise that is hardly
avoidable in any real quantum device. We also find that, although the Grover
driver Hamiltonian is potentially faster in the weak noise limit, the standard driver
Hamiltonian, which is actually more suitable to be implemented in existing quantum
hardware, results less sensitive to discrete noise.

\section*{Acknowledgments}
The authors thank Peter J. Love and Sergey Knysh for many useful discussions.
This work was supported by the Air Force Office of Scientific Research
under Grants FA9550-12-1-0046. \mbox{A.A.-G.} was supported by the National Science
Foundation under award CHE-1152291. \mbox{A.A.-G.} thanks the Corning Foundation
for their generous support.
The authors also acknowledge the Harvard Research Computing for the use of the
Odyssey cluster.

\section*{Author contributions}

S.M. and A.A.G. designed the research; S.M and G.G.G
designed and devised the software for the numerical analysis,
performed the research and analyzed the
data; S.M, G.G.G and A.A.G. wrote the paper.\\

\noindent\footnotesize{$^*$ These authors contributed equally to the paper.}

\normalsize

\section*{Appendix}
\label{sex:appendix}
\appendix

\section{Definition of ``computational time'' for an adiabatic quantum optimizer}
\label{sec:Tcomp}

In this appendix Section we provide a precise analysis of the concept of computational time required to achieve a solution of the problem at hand. In particular, we are interested to the case in which a larger probability of success may be achieved if the adiabatic quantum
optimizer is used for a shorter evolution time, but in repeated runs \cite{Boixo2014,Ronnow2014}.
This strategy trivially includes the possibility of performing a unique, long optimization run.

Let us define $p_S(T)$ as the probability of success of the adiabatic quantum optimizer
at fixed evolution time $T$. Recalling that the probability to (always) fail after $k$
attempts is given by $(1-p_S(T))^k$, the minimum number $K(T)$ of attempts to have a probability of $99\%$ to find the correct solution (at least once) results
\begin{equation}
	K(T) = \frac{\log(1-0.99)}{\log(1-p_S(T))} \, ,
\end{equation}
which leads to the definition of the computational time:
\begin{equation}\label{eq:Tcomp}
	T_\text{comp} = \min_T \Big\{ T \cdot K(T)\Big\} =
		\min_T \Big \{ T \cdot \frac{\log(1-0.99)}{\log(1-p_S(T))} \Big\}.
\end{equation}
As one can deduce from its definition, it is clear that $T_\text{comp} \leq T^*$,
where $T^*$ is the minimum evolution time to have $p_S(T^*) = 0.99$.\\


Consider now the case in which the quantum adiabatic optimizer has a hidden parameter
$q$, which is different from run to run and extracted from a distribution $p(q)$.
To give an explicit example, $q$ might take into account how the stochastic local noise
relates to the target state for the Grover problem, as described in
Section~\ref{sec:results}.
In this case, the probability of success of the quantum optimizer has to be averaged
over the possible values of the hidden parameter and becomes
\begin{equation}\label{eq:psucc_ave}
	\bar{p}_S(T) = \sum_q p(q)\, p_S(q,\,T),
\end{equation}
where $p_S(q,\,T)$ is the probability of success at fixed $q$. Consequently, the
computational time takes the form
\begin{equation}
	T_\text{comp} = \min_{T} \Big \{ T \cdot \frac{\log(1-0.99)}{\log(1-\bar{p}_S(T))} \Big\}.
\end{equation}

Assume that, for any given $q$, it is possible to exactly compute the spectral gap $g(s,\,q)$
at any time step $s$ of the adiabatic optimization. Therefore, an optimal schedule tailored
for that specific $q$ can be constructed, as described in \cite{Roland2002}, which has
an optimal evolution time given by
\begin{equation}
	T_\text{ann}(q) \propto \int_0^1 \frac{ds}{g^2(s,\,q)}\,.
\end{equation}
Since the calculation of the probability of success in Eq.~\eqref{eq:psucc_ave} requires the
evolution of the initial quantum state throughout the whole adiabatic calculation, we adopt two
main simplifications to avoid this extra overhead.
First, we assume that the optimal schedule obtained for a specific $q^*$ is not a good
adiabatic schedule for any other $q\neq q^*$, \emph{i.e.} that the probability
of any other $q\neq q^*$ is \emph{identically} zero
\begin{equation}\label{eq:psucc_simpl1}
	p_S(q,\,T|\,q^*) = \delta(q-q^*)p_S(q^*,\,T|\,q^*) \, .
\end{equation}
Second, we reduce the probability of success for $q^*$ to be a step function which is different
from zero only if $T > T_\text{ann}(q^*)$, namely
\begin{equation}\label{eq:psucc_simpl2}
	p_S(q^*,\,T|\,q^*) = \Theta(T - T_\text{ann}(q^*)) \, .
\end{equation}
Notice that both the above simplifications are quite conservative since we exclude the
possibility that an optimal schedule works (even partially!) for any other $q$ and that the
probability of success is strictly zero even for moderate evolution times.
Combining Eq.~\eqref{eq:psucc_simpl1} and Eq.~\eqref{eq:psucc_simpl2},
the computational time in Eq.~\eqref{eq:Tcomp} assumes the form
\begin{align}\label{eq:Tcomp2}
	T_\text{comp} &= \min_T \Big \{ T \cdot \frac{\log(1-0.99)}{\log(1-\bar p_S(T))} \Big\} \nonumber \\
		& = \min_{T,\,q^*}
			\Big \{ T \cdot \frac{\log(1-0.99)}{\log(1-p(q^*)\Theta(T-T_\text{ann}(q^*)))} \Big\}\nonumber \\
		& = \min_{q^*} \Big \{ T_\text{ann}(q^*) \cdot \frac{\log(1-0.99)}{\log(1-p(q^*)))} \Big\}.
\end{align}
It is important to notice that Eq.~\eqref{eq:Tcomp2} depends only on quantities
like $T_\text{ann}(q^*)$ and $p(q^*)$ which are properties of the model and not of the single
run. For example, for the Grover problem with noise in Section~\ref{sec:results},
both $T_\text{ann}(q^*)$ and $p(q^*)$ are completely determined by the noise model.

\section{``Tunneling'' model: Barrier around global minimum}
\label{sec:app_tunn_toy_model}

Here, we want to study a simple model that can be exactly solved using the exponential
reduction method. The peculiarity of this model is that the ground state of the problem
is surrounded by a ``high-energy barrier'' and, therefore, is hard to reach for a classical
simulated annealer. However, AQC might find the ground state very quickly due to the
tunneling effect as conjectured in Ref.~\cite{reichardt2004quantum,Farhi2002b}.
This example also demonstrates that our method can give rise to an exponential reduction
in the dimensionality even for problems where the gap behaves sub-exponentially, \emph{i.e.}
when the gap closes only polynomially.

Consider the simple problem Hamiltonian corresponding to the Hamming weight problem
\begin{equation}
	H_P = -\sum_{i=1}^n \hat\sigma_i^z \, ,
\end{equation}
which has a unique ground state, namely the configuration with all spins pointing up,
and a very simple energy landscape. The main idea is to add a barrier around this
unique ground state, that is to say we want to add a potential of the form
\begin{equation}
	V(\sigma) =
	\begin{cases}
		V_\alpha & \text{if }w(\sigma) = 1 \\
		0 & \text{otherwise},
	\end{cases}
\end{equation}
with $\alpha = \{1,\,\ldots,\,n\}$ the position of the single spin which is pointing down,
$V_\alpha > 0$ (but $V_\alpha < 0$ can be also used)
and $w(\cdot)$ the Hamming weight function. Let us define $\Ket{\alpha}$ as the state
in which all the spins are up apart from the $\alpha-$th spin which points down. Therefore,
the problem Hamiltonian becomes
\begin{equation}
	H_P = -\sum_{i=1}^n \hat\sigma_i^z + \sum_{\alpha=1}^n V_\alpha\KetBra{\alpha}{\alpha}.
\end{equation}
Using the standard driver Hamiltonian $H_D^{(S)} = -\sum_{i=1}^n \hat\sigma_i^x$, the AQO
Hamiltonian results
\begin{align}\label{eq:ham_toy_model}
	H_\text{AQO} &= -(1-s)\sum_{i=1}^n \hat\sigma_i^x
		-s \sum_{i=1}^n \hat\sigma_i^z + s \, \sum_{\alpha=1}^n V_\alpha\KetBra{\alpha}{\alpha} \nonumber \\
	  &= - \sqrt{s^2 + (1-s)^2} \, \sum_{i=1}^n H_i(s) + s \, \sum_{\alpha=1}^n V_\alpha\KetBra{\alpha}{\alpha} \\
	  &= - \sqrt{s^2 + (1-s)^2} \, \sum_{i=1}^n H_i(s) + s \, H_B \, , \nonumber
\end{align}
where $H_B$ is the barrier term and all $H_i(s)$ are identical single spin operators
which act on the $i-$th spin and whose explicit expression is given by:
\begin{align}
	H_i(s) &= \tfrac{1-s}{\sqrt{s^2 + (1-s)^2}} \hat\sigma_i^x +
              \tfrac{s  }{\sqrt{s^2 + (1-s)^2}} \hat\sigma_i^z \nonumber \\
	       &= \cos(\varphi_s) \hat\sigma_i^x + \sin(\varphi_s) \hat\sigma_i^z \, .
\end{align}
Since each $H_i(s)$ is a rotated Pauli matrix, it has eigenvalues $\pm 1$ with
corresponding eigenstates
\begin{align}
	\Ket{\phi_s^+} &= \frac{1}{\sqrt{2}} \left( \sqrt{1+\sin(\varphi_s)} \Ket{0} +
                                         \sqrt{1-\sin(\varphi_s)} \Ket{1} \right) \nonumber \\
                   &= \cos(\theta_s) \Ket{0} + \sin(\theta_s) \Ket{1} \, \nonumber \\
	\Ket{\phi_s^-} &= \frac{1}{\sqrt{2}} \left( \sqrt{1-\sin(\varphi_s)} \Ket{0} -
                                         \sqrt{1+\sin(\varphi_s)} \Ket{1} \right) \nonumber \\
                   &= \sin(\theta_s) \Ket{0} - \cos(\theta_s) \Ket{1} \, ,
\end{align}
with $\theta_s = \tfrac{\varphi_s}{2}$.
Let us call $\Ket{\phi^+_i(s)}$ and $\Ket{\phi^-_i(s)}$ the two eigenstates of $H_i(s)$
for any $s$. At this point, it is simple to understand that the Hamiltonian
$\sum_{i=1}^n H_i(s)$ has exactly $n+1$ energy levels characterized by the number
of $\Ket{\phi^+_i(s)}$ and $\Ket{\phi^-_i(s)}$ states in the product eigenstate.
In the $\{\Ket{\phi^\pm}\}$ basis, the states in the computational basis can be
written as
\begin{subequations}
\begin{equation}
	\Ket{0} = \cos(\theta_s)\Ket{\phi_s^+}+\sin(\theta_s)\Ket{\phi_s^-}
\end{equation}
and
\begin{equation}
	\Ket{1} = \sin(\theta_s)\Ket{\phi_s^+}-\cos(\theta_s)\Ket{\phi_s^-}.
\end{equation}
\end{subequations}
It is important to observe here that all the $\theta_s$ depend only on $s$
and not on the spin index $i$ since all local $H_i(s)$ are identical.
Interestingly, the Hamiltonian in Eq.~\eqref{eq:ham_toy_model} is the sum of two parts:
an Hamiltonian for which we know exactly the eigenenergies/eigenstates for any $s$ and
an Hamiltonian which non trivially acts only on $n$ states. Therefore, this model
can be exponentially reduces by using our method.

Before writing the explicit form of the overlap matrix, we introduce a simplified notation
in which $k(E)$ represent the number of $\Ket{\phi_s^-}$ states in each eigenvalue of
energy $E$ of $\sum_i H_i(s)$. With intuitive change of notation:
\begin{align}
    E(k)                    &=2k-n , \nonumber \\
    P_{\Omega_{E(k)}}       &=P_k  , \nonumber \\
    \mathcal{Z}_\alpha(E(k))&=\mathcal{Z}_\alpha(k) , \nonumber \\
    \Ket{E(k)_\alpha}       &=\Ket{k_\alpha} = \frac{P_k \Ket{\alpha}}{\mathcal{Z}_\alpha(k)} , \nonumber \\
    \lambda_{E(k)}          &=\lambda_k = \binom{n}{k} .
\end{align}
we can calculate
\begin{widetext}
\begin{align}
	\mathcal{Z}_\alpha(k)   &= \sqrt{\Braket{\alpha|P_{\Omega_E}|\alpha}} \nonumber \\
                            &= \sqrt{\tbinom{n}{k}} \sqrt{\tfrac{k  }{n} |\Braket{\phi_s^-|1}|^2 \, |\Braket{\phi_s^-|0}|^{2(k-1)} \, |\Braket{\phi_s^+|0}|^{2(n-k)}
                                   + \tfrac{n-k}{n} |\Braket{\phi_s^+|1}|^2 \, |\Braket{\phi_s^-|0}|^{2k}     \, |\Braket{\phi_s^+|0}|^{2(n-k-1)} } \nonumber \\
                            &= \sqrt{\tbinom{n}{k}} |\cos(\theta_s)|^{n-k} |\sin(\theta_s)|^{k} \sqrt{ \tfrac{k}{n} \tan^{-2}(\theta_s) + \tfrac{n-k}{n} \tan^2(\theta_s) }\\
   \Braket{k_\alpha|k^\prime _\beta} &= \delta_{k k^\prime}\, \frac{1}{\mathcal{Z}_\alpha(k) \mathcal{Z}_\beta(k)}
                                                \Braket{\alpha|P_k|\beta} \nonumber \\
                                      &= \delta_{k k^\prime}\, \mathcal{O}_{\alpha\beta}(k) \, .\label{eq:app_overlap}
\end{align}
\end{widetext}
The last line of Eq.~\eqref{eq:app_overlap} can be considered as the definition of the
overlap matrix $\mathcal{O}$ given in the basis $\left\{\ket{k_\alpha} \right\}_{\alpha,k}$.
The explicit expressions for the overlap matrix is (including the normalization):
\begin{widetext}
\begin{align}
    \mathcal{O}_{\alpha\alpha}(k) &= 1 \, , \nonumber \\
    \mathcal{O}_{\alpha\beta}(k)  &= \frac{1}{n-1} \frac{-2k(n-k) + k(k-1) \tan^{-2}(\theta_s) + (n-k)(n-k-1) \tan^2(\theta_s) }{k \tan^{-2}(\theta_s) + (n-k) \, , \tan^2(\theta_s)}
\end{align}
\end{widetext}
where, obviously, $\beta\neq\alpha$. Observe that the overlap element does not
directly depend on $\alpha,\beta$.

Finally, adopting the same notation as in Section~\ref{sec:orthogonalization}
we obtain the explicit form of both terms composing the reduced Hamiltonian
$H_{\text{eff}}$ in the basis $\{\Ket{\mathcal{E}_\mu}\}$ $\Big($notice that in our
simplified notation we have $\Ket{\mathcal{E}^\prime_\mu}\equiv\Ket{\mathcal{E}^{(E^\prime)}_\mu}$ $\Big)$:
\begin{widetext}
\begin{subequations}
\begin{align}
      \Braket{\mathcal{E}_\mu|H_B|\mathcal{E}^\prime_\nu} &=
                \sum_{\alpha=1}^n V_\alpha\Braket{\mathcal{E}_\mu|\alpha} \Braket{\alpha|\mathcal{E}^\prime_\nu}
             = \mathcal{Z}_\alpha(E)\,\mathcal{Z}_\alpha(E^\prime)
                \sum_{\alpha=1}^n V_\alpha\Braket{\mathcal{E}_\mu|E_\alpha} \Braket{E^\prime_\alpha|\mathcal{E}^\prime_\nu} \nonumber \\
             &= \mathcal{Z}_\alpha(E)\,\mathcal{Z}_\alpha(E^\prime) \left( \sum_{\alpha=1}^n V_\alpha\, T^{(E)}_{\mu \alpha} T^{(E^\prime)*}_{\nu \alpha} \right)
             = \mathcal{Z}_\alpha(E)\,\mathcal{Z}_\alpha(E^\prime) \left[ T^{(E)}T^{(E^\prime)\dagger} \right]_{\mu \nu} \, , \\
      \Braket{\mathcal{E}_\mu|\Sigma_i H_i(s)|\mathcal{E}^\prime_\nu} &=
                E \, \delta_{E E^\prime} \, \delta_{\mu \nu} \, .
\end{align}
\end{subequations}
\end{widetext}
Where $\alpha,\beta=1, \ldots, n$ are the indices corresponding to the states in
$H_B=\sum_{\alpha=1}^n V_\alpha\KetBra{\alpha}{\alpha}$,
while $\mu,\nu=1, \ldots, \kappa(E)$ are the indices labeling the orthonormal basis
states of the effective subspace of each $\Omega_E$ (\emph{i.e.} neglecting the state
spanning $\omega_E$).

\section{Grover Problem with local noise (Standard Driver)}
\label{sec:noisy_hamiltonian}

In this appendix Section, we will show how to apply our method for the Grover problem
in presence of local noise when the standard driver Hamiltonian is used. In the next
Section, we will show briefly the derivation when a Grover-style driver Hamiltonian
is used instead. Consider the following Grover problem Hamiltonian
\begin{equation}
	H_P = -n \KetBra{\omega}{\omega} + H_\text{dis},
\end{equation}
where $H_\text{dis} = \sum_{i=1}^n \epsilon_i \hat\sigma_i^z$ plays
the role of local disorder. If the standard
driver Hamiltonian is used, the AQO Hamiltonian results
\begin{equation}\label{eq:H_AQO_noise1}
	H_\text{AQO} = -s n \KetBra{\omega}{\omega} + s\sum_{i=1}^n \epsilon_i \hat\sigma_i^z - (1-s)\sum_{i=1}^n \hat\sigma_i^x.
\end{equation}
It is important to observe that the presence of the noise $\epsilon_i$ in
Eq.~\eqref{eq:H_AQO_noise1} breaks the spin-exchange symmetry. Therefore, no methods that
explicitly exploit that kind of symmetry can be used in this context. In the following, we
will show that the spectral gap of the Hamiltonian in Eq.~\eqref{eq:H_AQO_noise1} can be
calculated in a subspace whose dimension is exponentially reduced (compared to $2^n$),
even in presence of local disorder.
It is important to stress that our method allows the calculation of the spectral gap
\emph{without any perturbative expansion} around the small noise limit.

To begin with, let us apply a unitary transformation on Eq.~\eqref{eq:H_AQO_noise1} in
order to get rid of the sign of all $\epsilon_i$, namely
\begin{align}\label{eq:H_AQO_noise1}
	H^\prime_\text{AQO} &= -s n \KetBra{\omega^\prime}{\omega^\prime}
		- s\sum_{i=1}^n |\epsilon_i| \hat\sigma_i^z - (1-s)\sum_{i=1}^n \hat\sigma_i^x \nonumber \\
	&= -s n \KetBra{\omega^\prime}{\omega^\prime}
		- \sum_{i=1}^n \Big[s|\epsilon_i| \hat\sigma_i^z + (1-s)\hat\sigma_i^x\Big]\nonumber\\
	&= s\, H_B + \gamma(s)\, H_A.
\end{align}
where $\gamma(s) = \sqrt{(s|\epsilon|)^2 + (1-s)^2}$ and
\begin{subequations}
\begin{align}
	H_A &= - \sum_{i=1}^n \left[\frac{s|\epsilon_i|}{\gamma(s)} \hat\sigma_i^z + \frac{(1-s)}{\gamma(s)}\hat\sigma_i^x\right],\\
	H_B &= - n \KetBra{\omega^\prime}{\omega^\prime}.
\end{align}
\end{subequations}
Since we want to maintain the number of energy levels of $H_A$ polynomial in the number of
spins, we will consider the simple case where the noise is binomial, aka
$\epsilon_i= |\epsilon|\delta_i$ with $\delta_i = \pm 1$.
Observe that it is possible to have an exponential reduction even if the value of any
$|\epsilon_i|$ is chosen from a finite set of $p$ distinct (possibly incommensurable)
values: In this case, the number of energy levels $M$ of $H_A$ is upper bounded by
$M \leq (n+1)^p$.

\begin{figure*}[t!]
  \centering
  \includegraphics[width=0.95\textwidth]{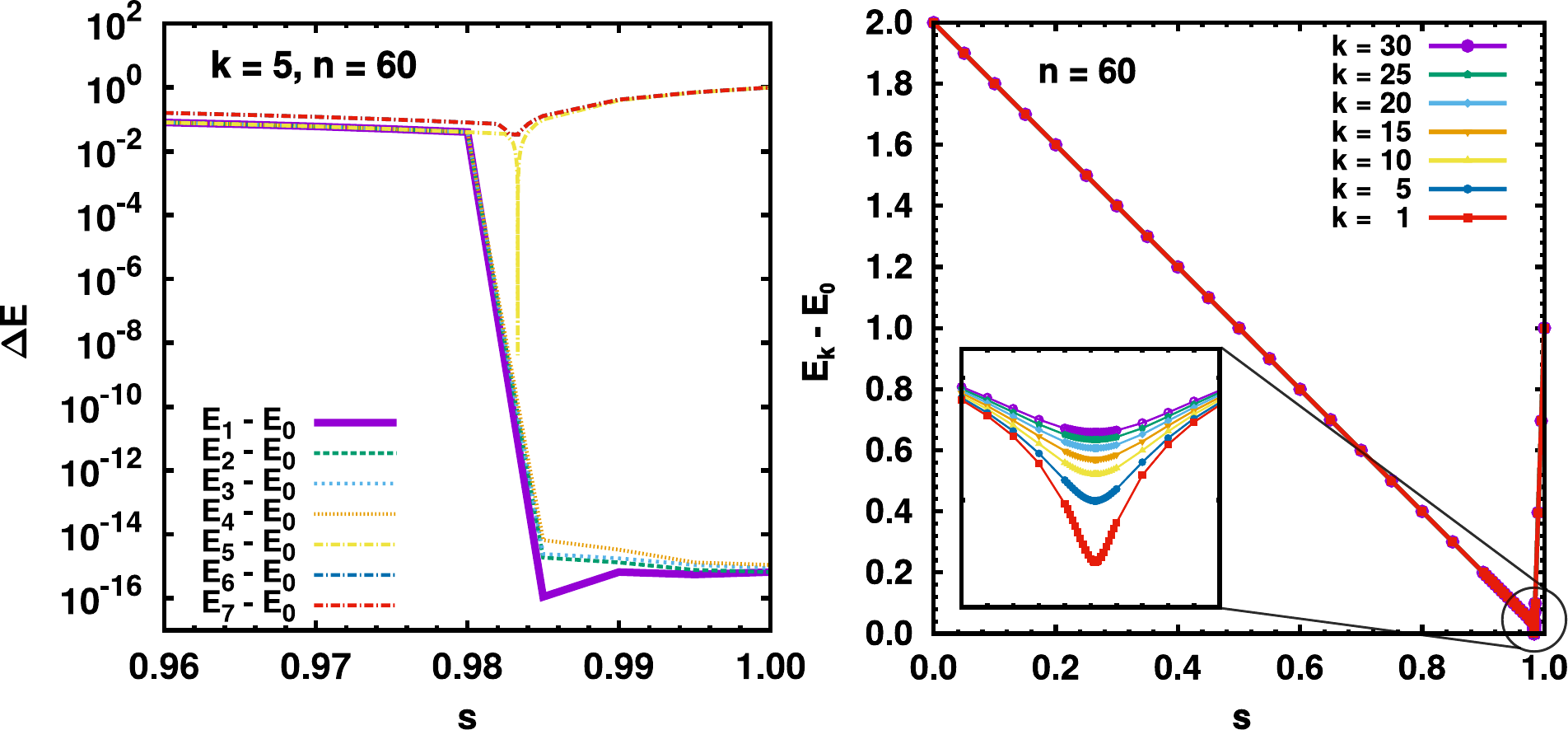}
  \caption{\label{fig:grover_multi_sol}
  	\textbf{Application of the proposed reduction method for the multi-solutions Grover problem.}
    {(Left panel) Difference in energy between the ground state and the $l-$th excited state, at fixed number of solutions $k = 5$
    and number of spins $n = 60$.
    (Right panel) Difference in energy between the ground state and the $k-$th excited state, by varying the number of solutions $k$ at
    fixed number of spins $n = 60$.}
  }
\end{figure*}

Therefore, the Hamiltonian $H_B$ in Eq.~\eqref{eq:H_AQO_noise1} becomes
\begin{equation}\label{eq:H_AQO_noise2}
H_A = - \sum_{i=1}^n \Big[\sin(\varphi_s)\,\hat\sigma_i^z + \cos(\varphi_s)\,\hat\sigma_i^x\Big]
    = -\sum_{i=1}^n \hat\sigma_i(s),
\end{equation}
with $\sin(\varphi_s) = \frac{s|\epsilon|}{\gamma(s)}$ and $\cos(\varphi_s) = \frac{(1-s)}{\gamma(s)}$.
As shown in Appendix~\ref{sec:app_tunn_toy_model}, all the $\hat\sigma_i(s)$ are identical
and their eigenstates (corresponding to the eigenvalues $\pm1$) are
\begin{subequations}\begin{align}
	\Ket{\phi_s^+} &= \frac{1}{\sqrt{2}} \left( \sqrt{1+\sin(\varphi_s)} \Ket{0} +
                                         \sqrt{1-\sin(\varphi_s)} \Ket{1} \right) \nonumber \\
                   &= \cos(\theta_s) \Ket{0} + \sin(\theta_s) \Ket{1},\\
	\Ket{\phi_s^-} &= \frac{1}{\sqrt{2}} \left( \sqrt{1-\sin(\varphi_s)} \Ket{0} -
                                         \sqrt{1+\sin(\varphi_s)} \Ket{1} \right) \nonumber \\
                   &= \sin(\theta_s) \Ket{0} - \cos(\theta_s) \Ket{1}.
\end{align}\end{subequations}
By inverting the above expressions, states in the computational basis can be expressed as
\begin{subequations}
\begin{align}
	\Ket{0} &= \cos(\theta_s)\Ket{\phi_s^+}+\sin(\theta_s)\Ket{\phi_s^-},\\
	\Ket{1} &= \sin(\theta_s)\Ket{\phi_s^+}-\cos(\theta_s)\Ket{\phi_s^-}.
\end{align}
\end{subequations}
Let us assume that $q = |\omega^\prime|$. Since the contribution $H_A$ to the Hamiltonian
$H^\prime_\text{AQO}$ as expressed in Eq.~\eqref{eq:H_AQO_noise1} is invariant by spin
exchange, we can always assume that all the spins in $\omega^\prime$ are ordered, namely
$\omega^\prime = \Ket{0\cdots01\cdots1}$ (but note that the total Hamiltonian
$H^\prime_\text{AQO}$ still violates the spin exchange symmetry!).
Using the same notation as introduced in Appendix~\ref{sec:app_tunn_toy_model}, we write
\begin{align}
    E(k)                    &=2k-n , \nonumber \\
    P_{\Omega_{E(k)}}       &=P_k  , \nonumber \\
    \mathcal{Z}_{E(k)}      &=\mathcal{Z}_k , \nonumber \\
    \Ket{E(k)}              &=\Ket{k} = \frac{P_k \Ket{\omega^\prime}}{\mathcal{Z}_k} , \nonumber \\
    \lambda_{E(k)}                &=\lambda_k = \binom{n}{k},
\end{align}
where $k$ is formally the number of $\Ket{\phi_s^-}$ in a given eigenstate of $H_B$
at energy $E$. $\lambda_k$ is the degeneracy of the energy level $E$.
Given the Hamiltonian in Eq.~\eqref{eq:H_AQO_noise2} and an arbitrary state in the computational base $\Ket{\omega^\prime}$, the normalization factor $\mathcal{Z}_k$ can
explicitly computed:
\begin{widetext}
\begin{equation}
	\mathcal{Z}^2_k = \Braket{\omega^\prime|P_k|\omega^\prime} =
		\sum_{l=0}^{\min\{k,\,q\}}\!\!\!\binom{q}{l}\binom{n-q}{k-l}
			\sin(\theta_s)^{2(q-l)+2(k-l)}\cos(\theta_s)^{2n-2(q-l)-2(k-l)}.
\end{equation}
\end{widetext}
Observe that if $\sin(\theta_s) = \cos(\theta_s) = \tfrac{1}{\sqrt{2}}$ (namely when
the disorder $|\epsilon_i|\to0$), the normalization factor becomes
$\mathcal{Z}_k = 2^{-n/2}\sqrt{\binom{n}{k}}$ for any choice $\omega^\prime$,
as expected.

\section{Grover Problem with local noise (Grover-style Driver)}
\label{sec:noisy_hamiltonian2}

In this appendix Section, we want to derive the exponential reduction for the Grover problem
in the presence of noise, when a Grover-style Hamiltonian is used instead of the standard
driver Hamiltonian (see Section~\ref{sec:noisy_hamiltonian}). Observe that since the
partitioning of the $H_{AQO}$ is different in the two cases, the final reduced Hamiltonian
have completely different forms. As in Section~\ref{sec:noisy_hamiltonian}, let us consider
the following Grover problem Hamiltonian
\begin{equation}
	H_P = -n \KetBra{\omega}{\omega} + H_\text{dis},
\end{equation}
where $H_\text{dis} = \sum_{i=1}^n \epsilon_i \hat\sigma_i^z$ plays the role of local disorder.
Adding the Grover-style driver Hamiltonian, the AQO Hamiltonian results
\begin{equation}\label{eq:H_AQO_noise3}
	H_\text{AQO} = -s n \KetBra{\omega}{\omega} + s\sum_{i=1}^n \epsilon_i \hat\sigma_i^z - (1-s)\KetBra{\psi_0}{\psi_0},
\end{equation}
where $\Ket{\psi_0}=\frac{1}{\sqrt{2^n}}\sum_z \Ket{z}$ is the equal superposition of all
the states in the computational basis.
After the application of an unitary transformation to get rid of all the sign of
$\epsilon_i$, the AQO Hamiltonian becomes
\begin{align}\label{eq:H_AQO_noise4}
	H^\prime_\text{AQO} &= -s n \KetBra{\omega^\prime}{\omega^\prime}
		- s\sum_{i=1}^n |\epsilon_i| \hat\sigma_i^z - (1-s) n \KetBra{\psi_0}{\psi_0}\nonumber\\
	&= n \Big(-s\KetBra{\omega^\prime}{\omega^\prime}- (1-s)\KetBra{\psi_0}{\psi_0}\Big)
		- s\sum_{i=1}^n |\epsilon_i| \hat\sigma_i^z \nonumber\\
	&= H_B + s\, H_A,
\end{align}
where
\begin{subequations}
\begin{align}
	H_A &= - \sum_{i=1}^n |\epsilon_i| \hat\sigma_i^z,\\
	H_B &= -n\Big(s\KetBra{\omega^\prime}{\omega^\prime}+ (1-s)\KetBra{\psi_0}{\psi_0}\Big).
\end{align}
\end{subequations}
As in Section~\ref{sec:noisy_hamiltonian}, we choose $\epsilon_i = |\epsilon|\delta_i$ with
$\delta_i = \pm1$. 
Therefore, $H_A$ assumes the simple form of a rescaled Hamming weight function, namely:
\begin{equation}
	H_A = -|\epsilon|\sum_{i=1}^n \hat\sigma^z_i.
\end{equation}
Once defined $E(k) = 2k-n$ and $P_k$ respectively the energy and the projector of the eigenspaces
of $H_A$, it is straightforward to follow Section~\ref{sec:orthogonalization} and identify the
relevant states in order to construct the reduced Hamiltonian:
\begin{subequations}
\begin{align}
	\Ket{E_k} &= \frac{P_k\Ket{\psi_0}}{\mathcal{Z}_k},\\
	\mathcal{Z}_k &= 2^{-n/2}\sqrt{\binom{n}{k}},\\
	\Ket{\omega^\prime} &= \alpha \Ket{E_q} + \sqrt{1-\alpha^2}\Ket{\mathcal{E}},
\end{align}
\end{subequations}
where $q$ is the Hamming weight of $\omega^\prime$, $\alpha = 1/\sqrt{\binom{n}{q}}$ and
$\Ket{\mathcal{E}}$ is an appropriate eigenstate which is orthogonal to $\Ket{E_q}$ and
lives in the $k-$th eigenspace of $H_A$.
Notice that the presence of $\Ket{\omega^\prime}$ in Eq.~\eqref{eq:H_AQO_noise4}
adds only one extra states (formally $\Ket{\mathcal{E}}$)
because $P_k\Ket{\omega^\prime}=\delta_{kq}\Ket{\omega^\prime}$, where $\delta_{kq}$
is the Kronecker delta.

\section{Grover problem with multiple solutions}
\label{sec:multi_sol_Grover}

In this appendix Section, we derive the exponential reduction for the Grover problem
when more solutions (\emph{i.e.} target states) are acceptable. As described in
Section~\ref{sec:Mk_ham}, the AQO Hamiltonian for the multi-solution Grover
problem using the standard driver Hamiltonian can be restricted to at most $k \times n$
orthogonal states, where $k$ and $n$ are, respectively, the number of solutions and the
number of spins composing the database register.\\

Let $\{\Ket{w_1},\,\ldots,\,\Ket{w_k}\}$ being the states representing the solutions
of the Grover problem. Their projections onto the eigenspaces of the standard driver
Hamiltonian can be written as
\begin{equation}
  \Ket{E_\alpha} = \frac{P_{\Omega_E}\Ket{w_\alpha}}{\mathcal{Z}_\alpha(E)}=\sqrt{1/\binom{n}{u(E)}}\sum_{x|w(x)=k}(-1)^{x\cdot w}\Ket{w},
\end{equation}
where $u(E) = (E+n)/2$ represents the number of spin up at a given energy $E$ of the
driver Hamiltonian, $x$ is an arbitrary bit configuration, and $w(x)$ is the Hamming
weight. After some combinatorial analysis, the overlap matrix $\Braket{E_\alpha|E_\beta}$
results to be:
\begin{equation}\label{eq:grover_multi_sol}
  \Braket{E_\alpha|E_\beta} = 1/\binom{n}{u(E)}\sum_{l=0}^{\min\{d_{\alpha\beta},\,u(E)\}}
    (-1)^l\binom{d_{\alpha\beta}}{l}\binom{n-d_{\alpha\beta}}{u(E)-l},
\end{equation}
where $d_{\alpha\beta}$ is the Hamming distance between $\omega_\alpha$ and $\omega_\beta$. As expected, if $d = 0$ the overlap is identically $1$, as
well as if $u(E) = 0$.

In Fig.~4, left panel, we show the difference in energy between
the ground state and the $l-$th excited state, at fixed number of solutions $k = 5$ and
number of spins $n = 60$, while in the right panel we show the difference in energy
between the ground state and the $k-$th excited state, by varying the number of solutions $k$
at fixed number of spins $n = 60$. As expected, the energy spectrum is $k-$degenerate
at $s = 1$, meaning that all the $k$ solutions belong to the reduced subspace obtained
through our method to exponentially reduced the effective dimensionality.\\ \\

\bibliographystyle{unsrt}
\bibliography{pra_exponential-reduction}{}

\end{document}